\documentclass[fleqn,10pt]{article}
\usepackage[utf8]{inputenc}
\usepackage[T1]{fontenc}
\usepackage{bm}
\usepackage{graphicx}
\usepackage{hyperref}

\title{Analysing the combined health, social and economic impacts of the corovanvirus pandemic using agent-based social simulation}
\author{Frank Dignum\\ Umeå University, Sweden
\and Virginia Dignum\\ Umeå University, Sweden
\and Paul Davidsson\\ Malmö University, Sweden
\and Amineh Ghorbani\\ TU Delft, NL
\and Mijke van der Hurk\\ Utrecht University, NL
\and Maarten Jensen\\ Umeå University, Sweden
\and Christian Kammler\\ Umeå University, Sweden
\and Fabian Lorig\\ Malmö University, Sweden
\and Luis Gustavo Ludescher\\ Umeå University, Sweden
\and Alexander Melchior\\ Utrecht University, NL
\and René Mellema\\ Umeå University, Sweden
\and Cezara Pastrav\\ Umeå University, Sweden
\and Loïs Vanhee\\University of Caen, France
\and Harko Verhagen\\Stockholm University, Sweden}
\begin{document}
\maketitle

\newpage



\section*{Analysing the combined health, social and economic impacts of the corovanvirus pandemic using agent-based social simulation}
\begin{abstract}
During the COVID-19 crisis there have been many difficult decisions governments and other decision makers had to make. E.g. do we go for a total lock down or keep schools open? How many people and which people should be tested? Although there are many good models from e.g. epidemiologists on the spread of the virus under certain conditions, these models do not directly translate into the interventions that can be taken by government. Neither can these models contribute to understand the economic and/or social consequences of the interventions. However, effective and sustainable solutions need to take into account this combination of factors. In this paper, we propose an agent-based social simulation tool, ASSOCC, that supports decision makers understand possible consequences of policy interventions, bu exploring the combined social, health and economic consequences of these interventions.
\end{abstract}

\maketitle

\section{Introduction}
In order to handle crises like the COVID-19 crisis, governments and decision makers at all levels are pressed to make decisions 
in a very short time span, based on very limited information \cite{rosenbaum2020facing}. Where Italy and later Spain are criticized by not being quick enough to react on the spread of the SARS-CoV-2 virus, one wonders whether others would have made different decisions at that time. Although health is considered of the prime importance, the interventions from governments are having huge economic impacts and possibly even bigger socio-psychological impact. Because the incubation time of the coronavirus is around two weeks it will take at least two weeks before the effects of any restriction on movements is visible. That is, decisions will everytime be two weeks behind the actual situation. Given this fact, many countries made a jump start and installed   possibly more severe restrictions than necessary in order to stay ahead of the effect. However, it also means that in many countries the restrictions will last for several more months. Extended lockdown and restrictions on movement are leading to social stress and the economic effects will have lasting consequences too. It can be expected that people will find unforeseen ways to circumvent the restrictions and there will be increased attempts to violate the restrictions, all potentially undoing the effects of the restriction. 

At the same time lockdowns are causing social stress, unemployment in several countries is rising at an incredible speed, causing even more social unrest. This is pushing the timing and type of strategies to exit lockdown and lessen the current restrictions, leading many governements to consider the introduction of tracking apps or other means to limit spread once restrictions are (partially) lifted, in order to limit the risks of a second or third pandemic wave. How and when should restrictions be lifted? Is it better to first re-start schools, or should transport and work take priority? What will be the effect of these strategies on the pandemic but also on the economy and social life of people and countries? 
It is also clear that interests and possibilities are not equal for all countries. Thus a preliminary lifting of restrictions in one country might adversely affect the spread in the neighbouring countries. Unless all cross border transport is halted, which is almost impossible given the economic and food dependencies between countries worldwide.

The above considerations make clear that health, social-psychological and economical perspectives are tightly coupled and all have a huge influence on the way the society copes with the COVID-19 crisis. Most of the time decision makers get their main advice from epidemiologists during pandemics. Other experts are sometimes added, but most models used by the experts focus on only one perspective, making it extremely hard to understand the combined effect of any policy measure across all aspects. 
This is especially serious if the effects of a restriction (or the lifting of a restriction) have an effect in one perspective that invalidates the assumptions made from another perspective \cite{benassy202013}. 
There is thus a need for ways to couple the different perspectives and model the interdependencies, such that these become visible and understandable. This will facilitate balanced decision making where all perspectives can be taken into account. Tools (like the one we propose) should thus facilitate the investigation of alternatives and highlight the fundamental choices to be made rather than trying to give one solution. 

In this paper, we present ASSOCC (Agent-based Social Simulation for the COVID-19 Crisis), an agent-based social simulation tool that supports decision makers gain insights on the possible effects of policies, by showing their interdependencies, and as such, making clear which are the underlying dilemmas that have to be addressed. Such understanding can lead to more acceptable solutions, adapted to the situation of each country and its 
current socio-economic state and that is sustainable from a long term perspective.

In the next section we will briefly discuss the current COVID-19 crisis situation. In section 3, we will discuss the different perspectives of the consequences of the crisis, and show what is needed to connect the different perspectives. In section 4, we describe the agent architecture at the core of the ASSOCC approach, which brings together the epidemiologic, social and economic perspectives, and show how this is implemented in a workable agent architecture that can be used in a social simulation framework. In section 5, we describe the practical application of the social simulation framework ASSOCC by exploring different example scenarios where the different perspectives are combined. Conclusions are drawn in section 6. 

\section{The coronavirus pandemic}
The COVID-19 crisis is characterized by very emotional debates and an atmosphere of crisis and panic \cite{khosravi2020}. When the pandemic spread from Asia to Europe it took some time to realise its possible consequences and what would be appropriate measures to prevent these consequences. Also the USA seemed to ignore what was happening in Asia and Europe for some time, causing a considerable delay in introducing defensive policies when the pandemic finally reached the country. In a country like the USA where little social security exists and government is not prepared to invest money in preparing for disasters the societal consequences are possibly even larger than in Europe \cite{hick2020}. This was also one of the lessons learned from the storm Katrina\cite{united2006}, but the lack of preparedeness still remains. In the USA alone already by end March 2020, over 6.6 million Americans had filed for unemployment\footnote{\url{https://www.theguardian.com/business/2020/apr/02/us-unemployment-coronavirus-economy}}. 

As the number of COVID-19 cases increased, policies had to be made quickly in order to prevent the rapid spread of the virus resulting in an overload of hospital and IC capacity. 
Given the lack of data about the coronavirus at the time of decision making, epidemic models based on those modeling earlier influenza epidemics were leading on making decisions. There was simply not enough information and not enough possibilities to take more focused measures that would target the right groups and still had the desired effect. The early days of the pandemic show a quick change of approaches as more became known about the coronavirus. E.g. the fact that in early stages it was believed that asymptomatic carriers were not able to transmit the virus, was influential in initial decisions concerning testing and contact tracing, and possibly to the high speed at which the virus spread initially. 

Currently many countries have introduced severe movement restrictions and full lockdown policies, with potentially huge social and economic consequences. However, also in these cases it is unclear what are the factors and motivations leading to the policy. In fact, an initial study by Oxford Univeristy shows that there is little correlation between the severity of the spread of the coronavirus and the stringency of the policies in place in different countries\footnote{\url{https://www.bsg.ox.ac.uk/research/research-projects/oxford-covid-19-government-response-tracker}}. 
The initial idea was that these measures would last for one or two months. However, at the moment there are already several countries speaking about a period of several months extending at least until mid 2020 or even longer. 

Based on data from previous pandemics, initial economic policies were based on the expectation of getting back to normal within a limited amount of time, with many governments sholdering the costs for the current period, it is increasingly clear that impact may be way above what governments can cope with, and a new `normal' economy will need to be found \cite{benassy202013}. 
And above that, the international dependencies of the world economy require that countries should coordinate their policies in order to sort maximum effect. A thing which is notoriously difficult and has not been improved by the recent attitude of the USA to go for its own interests first and show little solidarity with other countries.

From a sociological perspective there are not many theories that can be used to predict how the current situation will develop. However, some principles are clear. People have fundamental needs for affiliation and thus need to socialize. People can use the Internet for some of these needs, but physical proximity to other people is a basic need and cannot be witheld for a long period without possibly severe consequences. Keeping families locked up in their homes for long periods also will create problems of its own, even without considering the particular dangers for disfunctional families and domestic violence victims. New practices need to be formed where all members of the family will experience less privacy and autonomy and have to adjust their daily practices to accomodate the new situation. This is possible for short periods as tensions can be kept in reign. However, over long periods this might lead to conflicts and consequent problems of distress, increased domestic violence, suicides, etc. \cite{cluver2020}. These experiences might also affect families and society on the long term. People will change their behavior permanently to avoid similar situations. Thus e.g. people might be less inclined to travel, get close to other people, etc. The effects of these changes can be more subtle, but have a long lasting effect on the well being of society.

From the considerations above may be clear that policies impact epidemics, economics and society differently, and that a policy that may be beneficial from one perspective may lead to disastrous consequences from another perspective. As the crisis progresses and with it its impact increases, decision makers need to be aware of the complexity of the combined impact of policies. Means to support understanding this complexity are sorely needed as are tools that enable the design and analysis of many `what-if' scenarios and potential outcomes.

\section{Modeling complexity perspectives}
In this section, we describe the epidemics, economics and social science models that are needed to support decision makers on policies concerning the COVID-19 crisis and the complexity of combining these models.

\textit{Epidemiological} models are dominated by compartmental models, of which SIR~\cite{SEIR2018}, or SEIR, formulations are the most commonly used.
\begin{figure*}[h]
    \centering
    \includegraphics[width=80mm]{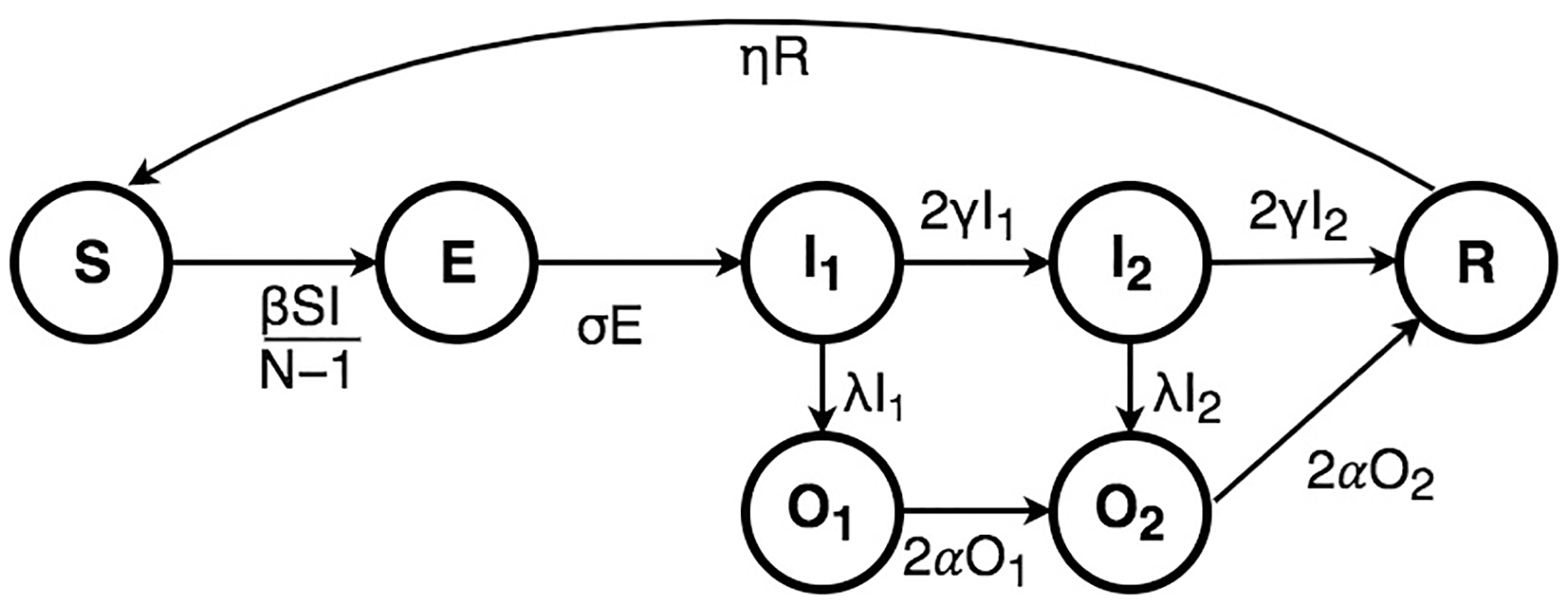}
    \caption{SEIR cycle as described in \cite{SEIR2018}}
    \label{fig:SEIR}
\end{figure*}
In compartmental models people are divided into compartments, as depicted in Figure \ref{fig:SEIR}.  SEIR shows how people start as being susceptible to a virus (S), then can become exposed (E) to the virus. From that state they can become infected ($I_1$). In that condition they can visit a doctor ($O_1$) or just stay home ($I_2$). From those states they can still visit a doctor and in the end they are either recovered (R) (or passed away). Once recovered people can become susceptible again if immunity does not last. The spread of the virus is determined by the probabilities with which persons move from one state to the next. Of course this figure gives a very simple picture of all complexities involved in epidemic models. However, what is clear from this picture is that people are a kind of passive containers of the virus that can infect others with a probability and can recover and become immune. There are no explicit actions that people undertake. A model like the above is often combined with a social network model that shows the speed of the spread along certain dimensions. When a person gets infected that is central in the network he will spread the virus in all directions and the epidemic spreads quick as well. 

These models are mathematical simplifications of infectious diseases, and allow for understanding how different situations may affect the outcome of the epidemic, by assuming that every person in the same compartment exhibit the same characteristics. That is, the model does not consider explicit human behaviour (as is also explained in the media nowadays\footnote{\url{https://www.vox.com/science-and-health/2020/4/10/21209961/coronavirus-models-covid-19-limitations-imhe}}) or consequences of interventions on actions of people \cite{heesterbeek2015modeling}. These are either considered outside of the model, or are transformed into an estimated effect on the parameters of the SEIR model that is uniform for all individuals in one compartment. E.g. the effect of closing schools on the spread of the corona virus can be interpreted as a lowering of factor $\delta$ in figure \ref{fig:SEIR} (meaning a lower number of places where the spread of the virus is possible and thus a lower probability for the virus to spread). However, what if the consequence of closing schools would be that children are staying with their grandparents or that they are brought together at the homes of other children rotating caring between families? 
In these cases, the number of places where the virus can spread may actually increase, which might outweigh the effect of closing the schools. It is possible to include all these factors in to the probability factor. However, by doing so, we loose the actual causal links between the different factors as these are not explicit part of the model and therefore cannot be easily identified and adjusted.\\

In \textit{economics}there are many competing models and theories. Without singling out a specific model, it can be said that in general 
economic models have difficulties in times of crisis \cite{kirman2010,colander2009financial}. Competing theories focus on different aspects of the economy and make different assumptions about the rest. The main issue all models struggle with (exactly like epidemiological models) is that of human behavior. Often, economic models take a `homo economicus' view on human behavior: a `rational' individual that always goes for maximum utility or profit in all circumstances. However, we all know this is not always true, as can be easily illustrated by the ultimatum game, an experimental economics game in which one player proposes how to divide a sum of money, e.g. 100 dollars, with the second party. If the second player rejects this division, neither gets anything. 
Rationally, the agent that gets offered should accept any offer as it is more than nothing (what he gets when refusing). However, empirical studies show that people only accept what they perceive as fair offers \cite{andersen2011stakes}. In our example, when more than 30-40 dollars are offered. So, fairness apparently also is worth something! It neatly illustrates that people have more motivations to take actions than mere utility of that action. Again, such motivations and values can be somehow incorporated in economic models, but such interpretation is not part of economic theory. I.e. economics does not directly provide an answer on how much fairness is worth. This will depend from the context and history. E.g. suppose an agent is pretty fair and offers the other agent 49 dollars and the other agent accepts. The next time the same agent offers the other agent 45, but now the other agent refuses due to the fact he feels he will get offered less and less. However, he might very well have accepted the 45 if it was offered the first time. 
The above is just one example to show that economic models are very good to explain and predict some behavior of people, but do not include all motivations and behaviors following from those.

The third perspective that is relevant for modeling of a pandemic is the \textit{social science}. In particular, social network analysis is often used to understand the possible ways a virus might spread \cite{firestone2011importance}. Nowadays much of the work in this area is related to online social media networks, because a lot of data is easily available on such networks. However, for the spread of a virus the physical social networks (between friends, family, colleagues, etc.) are those of main interest. Given that there is enough data to construct the physical social networks, they are a very good tool to determine which people might potentially be big virus spreaders. This can be due to their role (e.g. nurses in elderly care) or due to the fact that they have many contacts in different contexts (e.g. sports and culture) that are otherwise sparsely connected and they form bridges between densely connected communities. However, knowing which people or which roles one would like to target for controling the spread does not mean one can device effective policies. E.g. in the current pandemic it is clear that closing schools as being a possible bridge between communities is acceptable, while closing hospitals or elderly care centres is not. So, again, additional aspects have to be included in these models in order to use the theory in practice. More semantics for the nodes in the networks are needed as well as what the links between the nodes are constituted of. Do the links indicate the number of contacts per day? Can people also change the network based on their perception of a situation? I.e. avoid contacts, have different types of contacts, establish new contacts, etc. So, where social network analysis looks at people as nodes in a network, people are the ones that actually create, maintain and change the social network. When a government tries to contain the spread of a virus the social networks give a good indication where that might be done most effective, but how the people constituting the network will react to the policies is not included in the social network theory. Thus whether new links will arise bypassing previous links or other persons will take the place of so-called super spreaders cannot be derived from this theory.

\begin{figure*}[h]
    \centering
    \includegraphics[width=.6\linewidth]{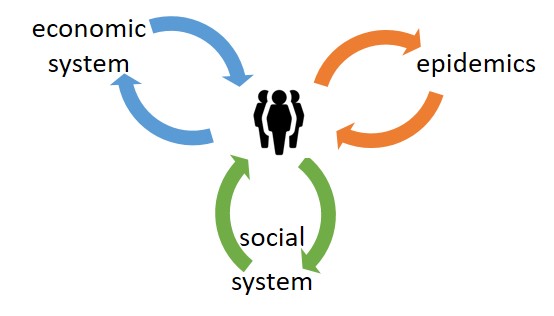}
    \caption{combining perspectives}
    \label{fig:perspectives}
\end{figure*}

Given the above short discussions of the different modeling perspectives of the crisis one can conclude that all perspectives include some assumptions about human behavior in their models. However, this behavior, and especially how people influence each other's behavior is not part of any of these models. We propose to take the human behavior as the central perspective and use it as a linking pin to connect the different perspectives as illustrated in Figure \ref{fig:perspectives}. 
In the next section we will discuss how our agent model from the ASSOCC framework can be used to fulfill this central role to couple the different perspectives.

\section{Agent Model}
\begin{figure*}[h]
    \centering
    \includegraphics[width=.7\linewidth]{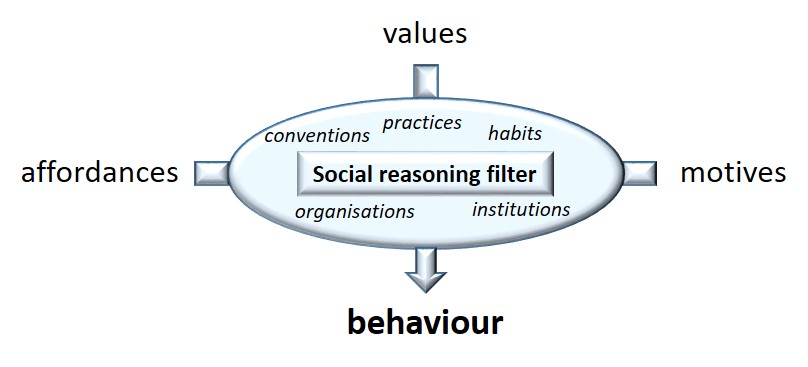}
    \caption{driving behaviour}
    \label{fig:behaviour}
\end{figure*}

The design of the ASSOCC framework is based on the fact that individuals always have to balance their needs over many contexts. In the research that we have done in the past twenty years we have come to the the sketch in Figure \ref{fig:behaviour} that illustrates how people manage this balancing act in their daily life.
We assume that people have a value system that is reasonably consistent both over times, contexts and domains. The value system is based on the Schwartz value circumplex \cite{schwartz_beyond_1994} that is quite universal. It depicts a number of basic values that everyone possesses and their relations. People differ in how they \textit{prioritize} values rather than on \textit{which} values they have. Although, priorities can differ between individuals they are reasonably consistent within cultural groups. Therefore the values can also be seen as linking individual drivers to the social group. We have used this abstract social science framework already in  \cite{vanhee_implementing_2011,vanhee_using_2015,Cranefield,cite:Heidari2018}. In order to use it in these simulations we have formalized and extended the framework such that it can be coupled to concrete behaviour preferences in each context.
Thus values give a stable guideline of behavior and they will be kept satisfied to a certain degree whenever possible. Thus, if “conservatism” is important to a person, she will, in general, direct her behavior to things that will benefit the preservation of the community.

The second type of drivers of behavior in figure \ref{fig:behaviour} are the motives that all people have in common.
This is based on the theory of McLelland \cite{mcclelland_human_1987}. The four basic motives that are distinguished are: {\emph achievement,affiliation,power and avoidance}. The achievement motive drives us to progress from the current situation to something better (whatever “better” might mean). The affiliation motive drives us to be together with other people and socialize. Thus we sometimes do things just to be doing it together with our friends and family. The power motive actually does not mean we want
power over others, but rather that we want to be autonomous. I.e. being able to do tasks without
anyone’s help. Finally, the avoidance motive lets us avoid situations in which we do not know how to behave or what to expect from others. Each of these motives is active all the time and whenever possible it will drive a concrete behavior. Thus, I might go to my grandparents to ask a question
rather than text them, just because I want to have a chat.

The third type of elements that determine behavior are the affordances that a context provides. These
affordances determine what kind of behavior is available and also what type of behavior is salient. E.g. in a bar one often drinks alcohol. Even though it is not obligatory it is salient and also afforded easily. Individuals have to balance between their values, their motives and the affordances to determine what behaviour would be more appropriate in each situation. As one can imagine this is quite tricky and will take too much time and energy if done in every situation from scratch. Therefore, in human society we have developed social structures to standardize situations and behaviors in order to package certain combinations that will be acceptable and usually good (even if not optimal). These social constructs are things like: norms, conventions, social practices, organizations, institutions. Note that these constructs give general guidelines or defaults of behavior, but are no physical restrictions on what is possible!

Implementing this whole architecture would be too inefficient for any social simulation. Therefore we use this as theoretical starting point, but translate it into a simpler model that is more efficient and scalable. Thus for each application we make choices which are the most important aspects of the above architecture and fix the resulting values and motives into a set of \emph{needs} as is illustrated in the following figure \ref{fig:needs}.
\begin{figure*}[h]
    \centering
    \includegraphics[width=.9\linewidth]{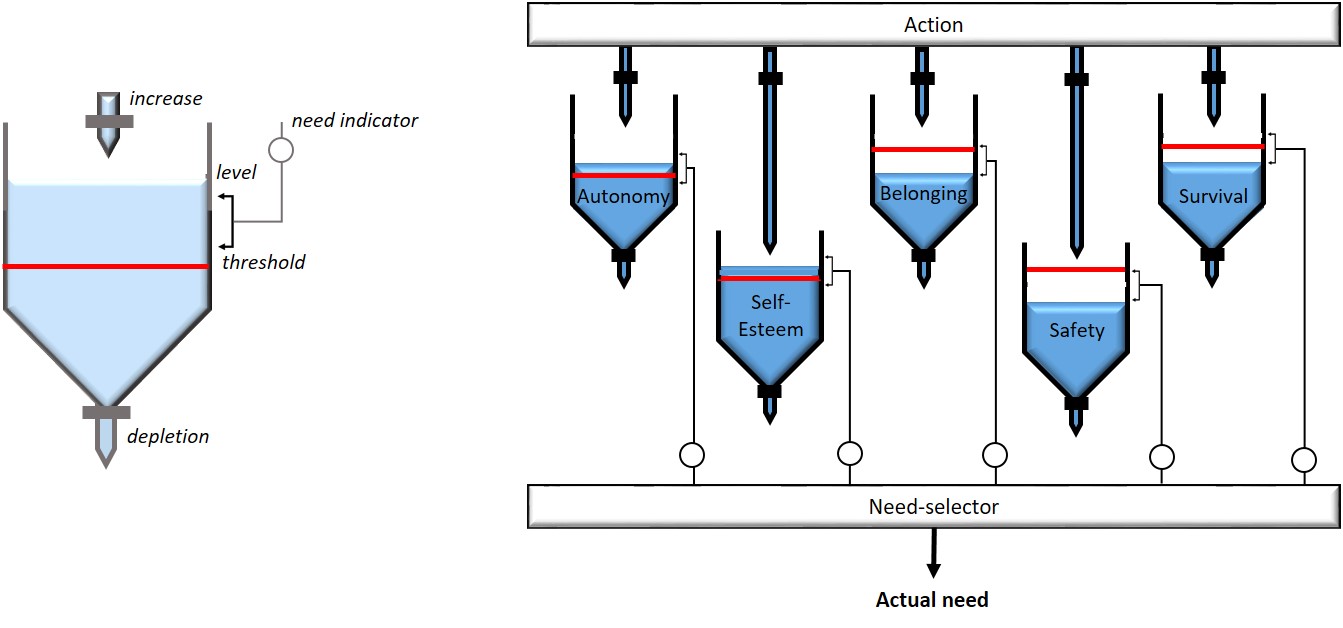}
    \caption{needs model}
    \label{fig:needs}
\end{figure*}

From the figure it can be seen that we model the values and motives as needs that deplete over time if nothing is done to satisfy them. The model prevents that an agent will only look at the need with the highest priority and only at other
ones when that need is completely satisfied. By calibrating the size and threshold and the depletion rate of each need we can calibrate and balance all the needs over a longer period, between different contexts and over several domains. E.g. using this model it becomes possible to decide for an individual whether it is more important to work a bit more or go home and be with the family. This simple model is the crux behind combining health, wealth and social wellbeing in a simulation model.

For our implementation of the COVID-19 crisis situation we have selected the following five needs that combine some values and motives that are more salient in the current crisis situation:
\begin{itemize}
    \item Safety
    \item Belonging
    \item Self-Esteem
    \item Autonomy
    \item Survival
\end{itemize}
From the four motives only the achievement motive is not included explicitly. Individual agents inherently are driven to take actions because they need to fulfill their needs. Longer term kinds of achievement, like getting educated or getting rich are not important for the purpose of this context. The avoidance motive is modeled through safety and survival, the affiliation is modeled through the need for belonging and the power motive is part of both the autonomy and self esteem. The needs also reflect the more salient values. The need for belonging models part of the value of universalism. The need for autonomy that for individual achievements. There is not a complete mapping as not all values are of importance given the type of activities and choices that the agents can make.\\ 
Some of the needs are subdivided. Safety has the more concrete needs of food-safety, financial-survival, risk-avoidance, compliance, and financial-safety. Food safety and financial survival represent that individuals have to have enough food and money to survive. Financial safety means that one has some buffer to pay other than the basic necessities. Compliance indicates whether complying to norms is important or not. Risk-avoidance in this context indicates whether people take actions that might get them infected or avoid any of those at all costs.\\
Given this core model of the agent decision model for behavior we can now describe the actual agent architecture as we have implemented it in ASSOCC.

\section{Architecture}
We have developed a NetLogo simulation consisting of a number (between 300 and 2500) of agents that exist in a grid. Agents can move, perceive other agents, and decide on their actions based on their individual characteristics and their perception of the environment. The environment constrains the physical actions of the agents but can also impose norms and regulations on their behavior. E.g. the agents must follow roads when moving between two places, but the environment can also describe rules of engagement such how many agents can occupy a certain location. Through interaction, agents can take over characteristics from the other agents, such as becoming infected with the coronavirus, or receive information. The main components of the simulation are:
\begin{itemize}
    \item \textbf{Agents}: representing individuals. Agents have needs and capabilities, but also personal characteristics such as risk aversion or the propensity to follow the law and recommendations from authorities. Needs include health, wealth and belonging. Capabilities indicate for instance their jobs or family situations. Agents need a minimum wealth value to survive which they receive by working or subsidies (or by living together with a working agent). In shops and workplaces, agents trade wealth for products and services. Agents pay tax to a central government that then uses this money for subsidies, and the maintenance of public services such as hospitals and schools.
    \item \textbf{Places}: representing homes, shops, hospitals, workplaces, schools, airports and stations. By assigning agents to homes, different households can be represented: families, students rooming together, retirement homes, three generation households and co-parenting divorced agents. The distribution of these households can be set in different combinations to analyse the situation in different cities or countries.
    \item \textbf{global functions}: under this heading we capture the general SEIR model of the corona virus which is used to give agents the status of infected, contagious, etc. This model also determines the contageousness of places like home, transport, shops, etc. based on a factor that represents fixed properties of a place (like size, time people spend there on average, whether it is indoor or outdoor) and density (how many people are there at the same time).\\
    Under this global functions we also capture economic rules that indicate tax and subsidies from government.\\
    Finally we also include the social networks and groups that exist under this heading. The social networks give information about \textit{normal} behavior and also provide clusters of agents performing activities together.
    \item \textbf{Policies}: describing interventions that can be taken by decision makers. For instance social distancing, testing or closing of schools and workplaces. Policies have complex effects for the health, wealth and well-being of all agents. Policies can be extended in many different ways to provide an experimentation environment for decision makers.
\end{itemize}

Agents can move between places and take the policies into consideration for their reasoning. As described in Section 4, agents' decisions are based on the container model of needs. 
These needs are satisfied by doing activities and decay over time. Needs may have different importance to each agent but the overall assumption is that agents will try to satisfy their most important need that is least satisfied at a given moment given the context. The context determines which choices are available at any given moment. Thus e.g. if agents have to work in a shop they will (normally) go to work even if the need for safety is high. But if they have work that can be done at home as well, they have a choice between going to work or staying home to work. In that case, their need for safety can make them decide to stay home.
Most needs are composite. For instance, safety is built up of food-safety, financial-survival, risk-avoidance, and compliance The first of these are most important for an agent's direct need to survive, so the safety need is defined as the minimum of the first two, and a weighted mean of the rest, where the weights are the importances assigned to each subneed. For example, the satisfation of food-safety is defined as having enough essential resources stocked up at home, such as food and medicine, measured over a two week period (i.e. the need is fully satisfied if the agent has enough supplies for the coming two weeks and decays from there). The only way to increase this need, is by going shopping for essential resources. However, going shopping, i.e. leaving the house may conflict with the need for safety, so the agent will need to balance these two needs in its decision to go shopping. Agents with a high level of risk-avoidance will be more likely to try to avoid the disease and thus want to stay away from large groups of people. We also assume that all agents have a certain need to comply with norms and regulations, which is satisfied by taking actions that conform to the rules, such as staying inside during lockdown, or going to school or work if that is requested from them. This need can have a negative value, in the case that the agent decides to break a rule.

The need for autonomy is satisfied when agents are able to follow their own plans. Agents satisfy their need for autonomy when they are able to make an “autonomous” decision. Lockdown policies block many of these actions, which means that when an agent reaches a too low level of autonomy it may decide to break lockdown. However, to regulate this effect and not provide agents with too strong incentives to break the lockdown, the 'compliance' is used as a regulating factor.
Finally, the need for survival represents an agent’s need to rest if it is sick. This need can be satisfied by resting at home if the agent believes it is sick, and depletes if it does anything else while it believes to be sick. Under this need we also fitted the conformity need. People will conform to what other people do if they are uncertain about the context and which action is the best to take. Conforming to others is safe as it is usually good to do what others do. Thus it contributes to (social) survival.

\begin{figure*}
    \centering
    \includegraphics[width=100mm]{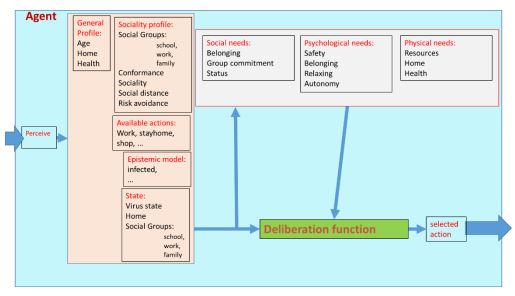}
    \caption{ASSOCC agent Architecture}
    \label{fig:architecture}
\end{figure*}

As can be seen above, agents in ASSOCC have a wide range of needs. They range over the social, health and economic dimensions of society and can therefore also show how interventions intended to remedy e.g. the spread of the virus can influence other dimensions and due to these dependencies do not have the intended effect. In the next section we describe some examples of scenarios we have simulated in ASSOCC to illustrate the use, the range of possible scenarios and importance of the approach.

\section{Example scenarios}

In this section we briefly discuss two scenarios we have simulated with ASSOCC. They are about quite different aspects of the COVID-19 crisis. The first is on the question whether schools should be closed and employees work at home during the pandemic. The second investigates some economic effects of having a lockdown for several weeks or even months. Both scenarios are built on the same model as described above, which shows the richness of the model and possibilities for exploration of new policies.
\subsection{Schools and Working at Home}
Typically, schools are places where both children and parents from a community gather, which potentially leads to spreading of the virus. When epidemics emerge, one common measure to be taken is the closure of schools. In case of influenza-like illnesses it has been proven that this is an effective measure, since children are highly susceptible for that disease. However, the COVID-19 pandemic is different, as it seems to be less contageous for children.\\
Sweden, as one of the few European countries, has kept primary and junior school (0-16 years old) open, based on the premise that closing schools would mean many parents with jobs that are vital to society, like healthcare personnel, would have to stay home and take care of their children instead of going to work. This would delay efforts to stem the spread of the virus. In \cite{Chang2020} it is already indicated that closing schools would not have much effect in the Australian situation. In this scenario, we explore potential consequences of keeping schools open on the spread of the virus but also on social and economic aspects. We model the direct and indirect effect on the spread of the virus when schools are closed and people work from home.

We assume that schools will be closed as soon as a certain amount of infected people within the city has been reached. Different thresholds have been tested, but in this paper schools are closed whenever one infected person is detected.\\
The scenario assumes that when children are staying at home, at least one adult should be at home to take care of them. This caregiver is assumed to work from home for the duration of the school closure. We will in particular look at the effect on the spreading of the coronavirus when schools are closed and thus parents working at home compared to parents already working at home with children still going to school. 

\subsubsection{Results}
\begin{figure}[h] \label{fig:schoolresults} 
  \begin{minipage}[b]{0.5\linewidth}
    \centering
    \includegraphics[width=.9\linewidth]{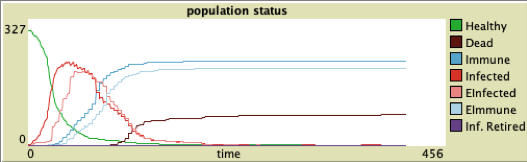} 
    a:~closing schools (health)
  \end{minipage} 
  \begin{minipage}[b]{0.5\linewidth}
    \centering
    \includegraphics[width=.9\linewidth]{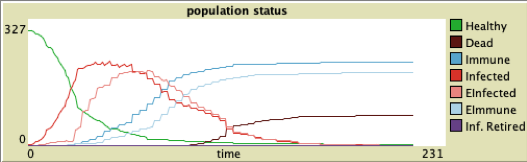} 
    b:~working at home (health)
  \end{minipage} 
  \begin{minipage}[b]{0.5\linewidth}
    \centering
    \includegraphics[width=.9\linewidth]{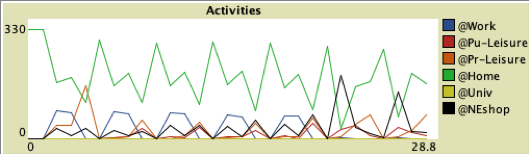} 
    c:~closing schools (activity) 
  \end{minipage}
  \hfill
  \begin{minipage}[b]{0.5\linewidth}
    \centering
    \includegraphics[width=.9\linewidth]{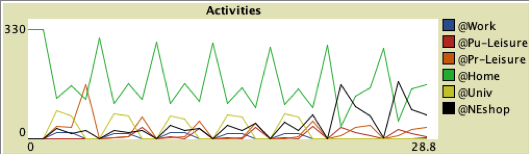} 
    d:~working at home (activity) 
  \end{minipage} 
  \caption{results}
\end{figure}

In figure \ref{fig:schoolresults} the results of closing the schools and working at home are shown. We show a graph with the health status of the population and the activities taking place. When only the schools are closed (the left two graphs), some people are forced to work at home too, as they have to take care of their children. 

The scale on the horizontal axis denotes time. Four ticks make up one day. So, the runs cover about 2-4 months (depending on whether any differences still appear after 2 months). The vertical axes show the number of agents which is around 330. This number can vary slightly as it depends on the household distribution and the way the households are generated. The results are based on 40 runs.\\ 
Although the number of people at home is of course much higher than people in other places (the green line in the bottom plots) compared to the baseline simulation, which implies less spreading of the virus, the peak of infected people is higher (around 210 and 215 resp.). The peak is reached around the same time, i.e. 40 ticks (or 10 days). The number of people that died is 80 and 78 resp. This means that the differences are not significant. 

We also see a difference in people going to non-essential shops (the black line in the activity graphs). The number of people at non-essential shops, on the right where the last two peaks (Saturday and Sunday) show that people go out shopping in the weekend. 

The above findings show that the measures do not lower the peak as expected, but even increase that peak. This can be explained by the fact that more people want to leave the house during the weekend, as they have worked at home during the week with less social interaction. At these public places typically people of different communities are gathered, while at work or school, more or less the same group of people are present. Thus closing the schools only increases the spreading of the virus. Thus without additional restrictions, like social distancing and closing restaurants, the effect of working at home or closing schools is surpassed by the side effect, namely people wanting to go out.

\subsection{Economic Effects}
The COVID-19 pandemic has severe medical and social consequences. But it has vast economic consequences as well. A recession is expected, and predictions show that it can be way more severe than the banking crisis (2009) recession.
Thus governments are trying to minimize the economic impact of the pandemic and try to minimize the “financial issues” people and companies have by giving massive financial support. Also, once the pandemic has been dealt with, “restarting” the economy is a big challenge as many people have lost their jobs and economic activity has slowed to a minimum. With this scenario we demonstrate the relation between the pandemic, health measures and the economy. This shows the complex and interconnected nature of the situation.

In many countries the government has locked down the country and closed down business operations.This cuts the cycle of income of companies and subseqently either people become unemployed or governments temporarily take over paying the wages of the employees. In this scenario we show two different situations. The first shows what happens with a lockdown and no government support, the second shows the situation when government supports companies by taking over wages.

When government closes all non-essential shops to prevent spread and orders a lockdown the curve of the pandemic is really flattened. Within the time span of our simulation we see that we are still not experiencing many infections and hardly any deaths. Note that lockdown here starts when only one person was infected. So, much earlier than in real life!
People are almost all working at home.
The essential shops seem to be doing well. This can be seen as hoarding behavior of people that want to make sure they have enough essential products and have still money to buy things. They do not buy anything non-essential as those shops are all closed (and we have no on-line shopping in our simulation!). The non-essential shops do not go bankrupt in the simulation because we did assume here that they could postpone all fixed costs, like loans, rent, etc. If these fixed costs are taking into account all the non-essential shops will go bankrupt pretty quick without government intervention. The velocity of the money flowing through the system decreases due to the lockdown. 

\begin{figure}[h] \label{fig:lockdownnosubsidy} 
  \begin{minipage}[b]{0.5\linewidth}
    \centering
    \includegraphics[width=.9\linewidth]{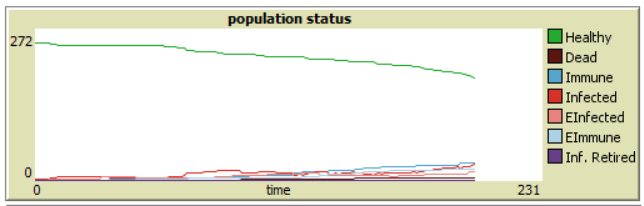} 
    a:~health of population
  \end{minipage} 
  \begin{minipage}[b]{0.5\linewidth}
    \centering
    \includegraphics[width=.9\linewidth]{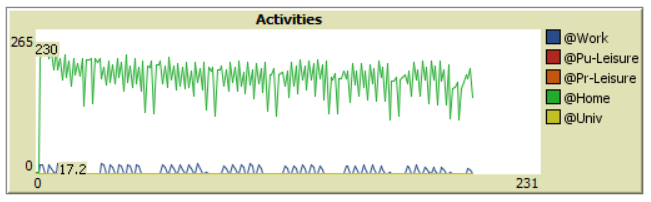} 
    b:~activity
  \end{minipage} 
  \begin{minipage}[b]{0.5\linewidth}
    \centering
    \includegraphics[width=.9\linewidth]{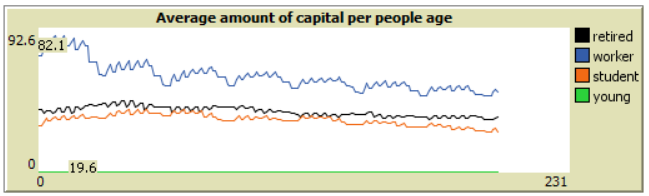} 
    c:~wealth by age group 
  \end{minipage}
  \hfill
  \begin{minipage}[b]{0.5\linewidth}
    \centering
    \includegraphics[width=.9\linewidth]{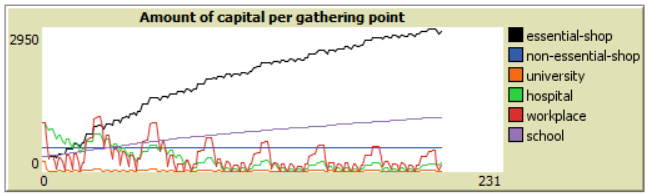} 
    d:~wealth per organization 
  \end{minipage} 
  \caption{results no subsidy}
\end{figure}

In the second situation government takes over wages from all people working in the non-essential shops. Overall we see capital increasing slightly. This can be explained due to the fact that people cannot spend money on non-essential products and leisure activities (going to cafes, sport events, etc.) while the number of unemployed/non-payed people is relatively small.

\begin{figure}[h] \label{fig:lockdownsubsidy} 
  \begin{minipage}[b]{0.5\linewidth}
    \centering
    \includegraphics[width=.9\linewidth]{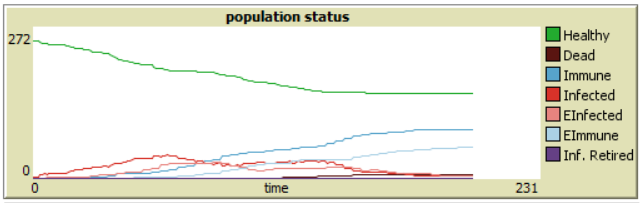} 
    a:~health of population
  \end{minipage} 
  \begin{minipage}[b]{0.5\linewidth}
    \centering
    \includegraphics[width=.9\linewidth]{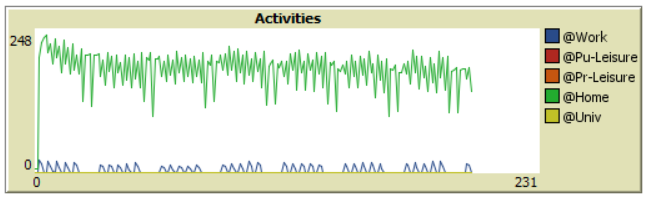} 
    b:~activity
  \end{minipage} 
  \begin{minipage}[b]{0.5\linewidth}
    \centering
    \includegraphics[width=.9\linewidth]{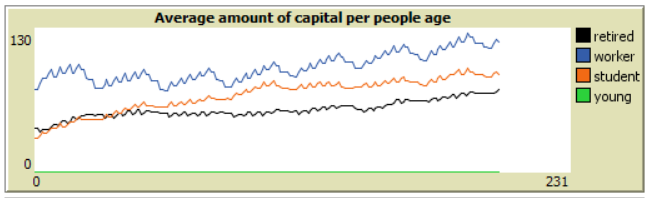} 
    c:~capital per age group 
  \end{minipage}
  \hfill
  \begin{minipage}[b]{0.5\linewidth}
    \centering
    \includegraphics[width=.9\linewidth]{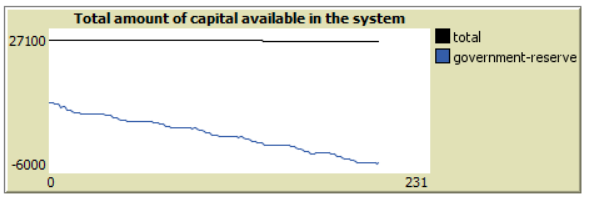} 
    d:~government reserves 
  \end{minipage} 
  \caption{results with government subsidy}
\end{figure}

The economic activity is kept up by the government. However, the government reserves are quickly depleting as less tax is coming in and more subsidies are paid. This situation is not sustainable!

\section{Conclusions}
In this paper we have shown that in crises like the COVID-19 crisis the interventions of government have to be made quick and are often based on too little information and only partial insights of the consequences of these interventions. In most cases the health perspective is the leading perspective to inform government decisions. However, the models used by epidemiologists lack the part on human behavior. Therefore they have to translate government interventions and the expected reactions from citizens to parameters into parameters of the epidemic model without being able to check for interdependencies. We have shown how the ASSOCC platform can be a good addition to current epidemiological and economic models by putting the human behavior model central and from that core connect the health, wealth and social perspectives.\\
In general there are several advantages of using an agent based tool like ASSOCC. First a tool like ASSOCC explicitly avoids providing detailed predictions, but rather supports investigating the consequences in all perspectives from government policies. By comparing variations of policies (like government subsidizing people or not in the economic scenario) it becomes clear what are the consequences and which are the fundamental choices that have to be made. Taking this stance avoids politicians being able to just blame the model for giving a wrong advice. Technology should not give single solutions to decision makers, but should support the decision makers to have a good insight in the consequences of their decisions and which are the fundamental priorities that have to be weighed. E.g. the joy of life of elderly people against the risks of meeting their family and becoming infected. Or the risk of contagion happening through schools vs. the social disruption when schools are closed.\\
Due to the agent based nature it is also possible to explain where results come from. E.g. closing schools might not have any positive effect on the spread of the virus due to all kinds of side effects. In the scenario we could see that people were going more to non-essential shops in the weekend because they had to be more at home to take care of the children during the week. Their need for belonging went up and going to the shops was the only option left as other leisure places were closed. Tracing this behavior back in this way helps to explain what is the basic cause and where in the chain something can be done if one wants to avoid this behavior. Each of the steps in these causal chains is based on a solid theory (social-psychological, economic or epidemiological in our case). And thus can be checked both on its plausability as well as on its theoretical foundation.\\
Using the agent based models makes it easier to adjust parameters in the simulation based on differences in demographics and cultural dimensions in different countries, but also in different regions or even neighbourhoods in big towns. Therefore more finegrained analysis can be performed on the impact of government interventions on different parts of the country. E.g. rich and poor neighbourhoods or urban and rural areas can be affected in different ways by the same measure.\\
Finally, we should stress that we do not claim that agent based models like used in ASSOCC should replace domain specific models! Actually one could use ASSOCC to simulate intervention scenarios to study the human reactions to it and the consequences. Having a good insight in these dependencies one can feed the domain specific models with better information to make precize optimizations or predictions for the effect of that intervention. Thus in this way the strength of the different types of models can be combined rather than seen as competing.

\bibliographystyle{abbrv}
\bibliography{large}



\noindent\textbf{Supplementary information}
The full project is available at \url{https://simassocc.org}. The NetLogo and GUI code is available at \url{https://github.com/lvanhee/COVID-sim} including a few runnable scenarios, described in the website.

\end{document}